\documentclass[review,sort&compress,number]{elsarticle}

\usepackage[T1]{fontenc}
\usepackage[utf8]{inputenc}
\usepackage{lmodern}
\usepackage{microtype}
\usepackage[english]{babel}
\usepackage{enumitem}
\usepackage{amsmath}
\usepackage{amssymb}
\usepackage{graphicx}
\usepackage{wrapfig}
\usepackage{float}
\usepackage{xcolor}
\usepackage{algorithm}
\usepackage{algorithmic}
\usepackage{multirow}
\usepackage{bbm}
\usepackage{gensymb}
\usepackage{textcomp}
\usepackage{gensymb}
\usepackage{lineno}
\usepackage{color, soul}

\usepackage{fancyhdr}
\begin{document}

\begin{frontmatter}
\title{Investigating the time dynamics of wind speed in complex terrains by using the Fisher-Shannon method}
\author{Fabian Guignard$^{1}$, Michele Lovallo$^{2}$, Mohamed Laib$^{1}$, Jean Golay$^{1}$, Mikhail Kanevski$^1$,Nora Helbig$^3$, Luciano Telesca$^4$ }
\address{$^1$IDYST, Faculty of Geosciences and Environment, University of Lausanne, Switzerland. \\
$^2$Agenzia Regionale per la Protezione dell' Ambiente di Basilicata, Potenza, Italy \\
$^3$ WSL Institute for Snow and Avalanche Research SLF, Davos, Switzerland\\
$^4$CNR, Istituto di Metodologie per l’Analisi Ambientale, Tito (PZ), Italy \\
Corresponding author: Fabian.Guignard@unil.ch}

\begin{abstract}

In this paper, the time dynamics of the daily means of wind speed measured in complex mountainous regions are  investigated. 
For 293 measuring stations distributed over all Switzerland,  the Fisher information measure and the Shannon entropy power are calculated. 
The results reveal a clear relationship between the computed measures and both the elevation of the wind stations and the slope of the measuring sites. 
In particular, the Shannon entropy power and the Fisher information measure have their highest (respectively lowest) values in the Alps, where the time dynamics of wind speed follows a more disordered pattern.
The spatial mapping of the calculated quantities allows the identification of two regions, which is in agreement with the topography of the Swiss territory. The present study could contribute to a better characterization of the temporal dynamics of wind speed in complex mountainous terrain.

\end{abstract}


\end{frontmatter}

\section{Introduction}
\label{intro}

Wind power is widely produced with little environmental pollution \cite{Nematollahi2016}. In fact, it has been gaining a growing attention in the renewable energy context, because it substitutes conventional fuels, playing thus a major role in the future energy production \cite{Edenhofer2012}. Wind power can be efficiently produced (the total wind capacity of the world increased by more than 10\% between 2013 and 2015 \cite{Dai2017}) with a modest environmental impact and  becomes very competitive from an economic viewpoint \cite{Nematollahi2016}.

In mountainous regions like the Alps, the wind speed is highly variable  in time as well as in space. Topography significantly influences the time dynamics of wind speed \cite{Barry1992, Mortensen1998}. Orographic features like ridge crests, deep valleys, or other rough landscapes can induce changes on boundary layer flows \cite{Etienne2010}. The Alps are featured by several locally varying climatic phenomena, natural channeling effects, and thermally induced circulations that characterize, for instance, wind speed as very high at one location but very low in a near valley, making the spatial interpolation of wind speed an   arduous task due to the large variability within small areas \cite{Tveito2008}. The investigation of wind speed within the atmospheric boundary layer is always challenging, because its large fluctuations and non-linearity make the space-time variability high \cite{Etienne2010}.
Nevertheless, recently the influence of topography on surface wind speed was investigated and a subgrid parameterization of the unresolved topographic impact was developed together with a statistical downscaling method of coarse-scale wind speed to finer scales \cite{Helbig2017}. Additionally, modeling wind speed and its regionalization is not easy \cite{Goyette2001}, mainly because many turbulence effects and roughness factors are present \cite{Stull1988}. 

Several investigations on wind speed measured over the territory of Switzerland have been carried out \cite{Schaffner2006, Jungo2002}. Etienne et al. \cite{Etienne2010} applied the Generalized Additive Models (GAMs) to regionalize wind speeds recorded at Swiss weather stations by using physiographic parameters. They provided reliable wind predictions based on the 98th percentile of the daily maximum wind speed, and found that wind speed depends on elevation and roughness of the mountains. Jungo et al. \cite{Jungo2002} applied the Principal Component Analysis (PCA) and the Cluster Analysis (CA) to several Swiss meteorological stations, obtaining clusters of stations, with the spatial distribution depending on the complexity of terrain. Weber and Furger \cite{Weber2001} performed an automated classification algorithm for one year wind data and found sixteen different near-surface wind flow patterns, which can have complex structures, like large-scale winds and locally forced wind systems interplay. Robert et al. \cite{Robert} applied General Regression Neural Networks (GRNN) to interpolate monthly wind speed in complex Alpine orography, using some Swiss meteorological network data as training data to obtain a  relationship between topography and wind speed. Telesca et al. \cite{Telesca2016} investigated the power spectrum and multifractality of several wind speed time series in Switzerland, finding the presence of periodicities of 24 hours  and 12 hours, related to the daily cycle of temperature and pressure.

In this study, time series of daily mean of wind speed  recorded by two  monitoring networks in Switzerland are analyzed: the SwissMetNet network (from MeteoSwiss), which covers the entire territory of Switzerland, and the IMIS network  (from WSL Institute for Snow and Avalanche Research SLF), covering very densely the Alps mountains. The large variability of topographic characteristics of the Swiss territory might induce a complex time dynamics in wind speed. Therefore, in order to quantify such complexity in  wind speed time series measured in Switzerland, the Fisher-Shannon (FS) method is employed \cite{MARTIN1999}. 




\section{Data and exploratory analysis}
\label{sec:1}
\begin{figure}
\centering
\includegraphics[width=\linewidth]{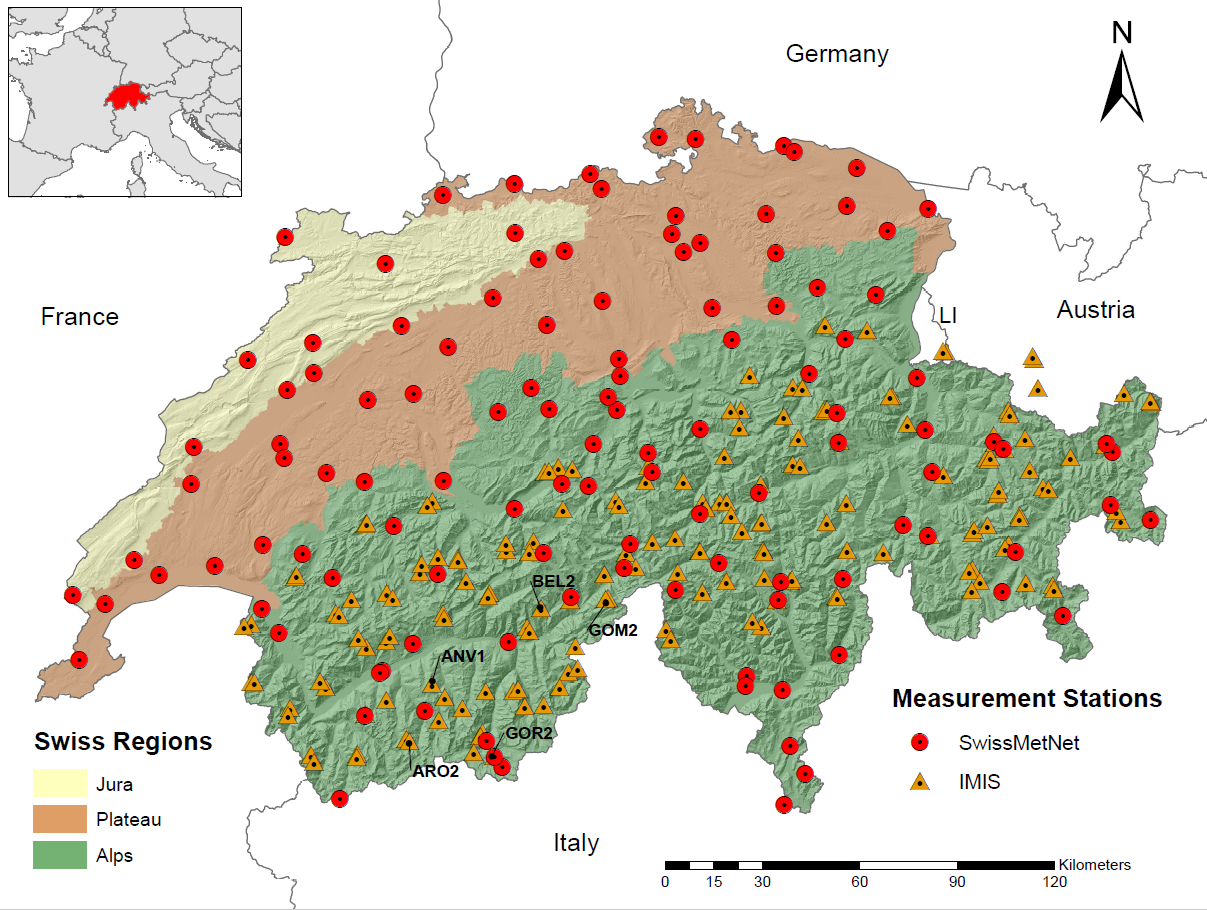}
\caption{Study area delimiting the three Swiss regions, adapted from \cite{Carmen2015}, as well as the locations of wind measurement stations.}
\label{fig1}  
\end{figure}

In this paper the time dynamics of wind speed measured from 2012 to 2016 by two networks in Switzerland are investigated. The first network (SwissMetNet), covering rather homogeneously all Swiss territory at various elevations,  consists of 119 stations that measure wind speed at 10-min sampling time and is managed by the Federal Office of Meteorology and Climatology of Switzerland (MeteoSwiss). The second network (IMIS), covering very densely the Swiss Alps, consists of  174 stations that measure wind speed at 30-min sampling time and is managed by the WSL Institute for Snow and Avalanche Research SLF. Fig. \ref{fig1} shows the locations of the wind stations and the three Swiss regions (Jura, Plateau and Alps) delimited by SwissTopo (Swiss Federal Office of Topography) based on geological and geomorphological features (adapted from \cite{Carmen2015}).

\begin{figure}
\centering
\includegraphics[width=\linewidth]{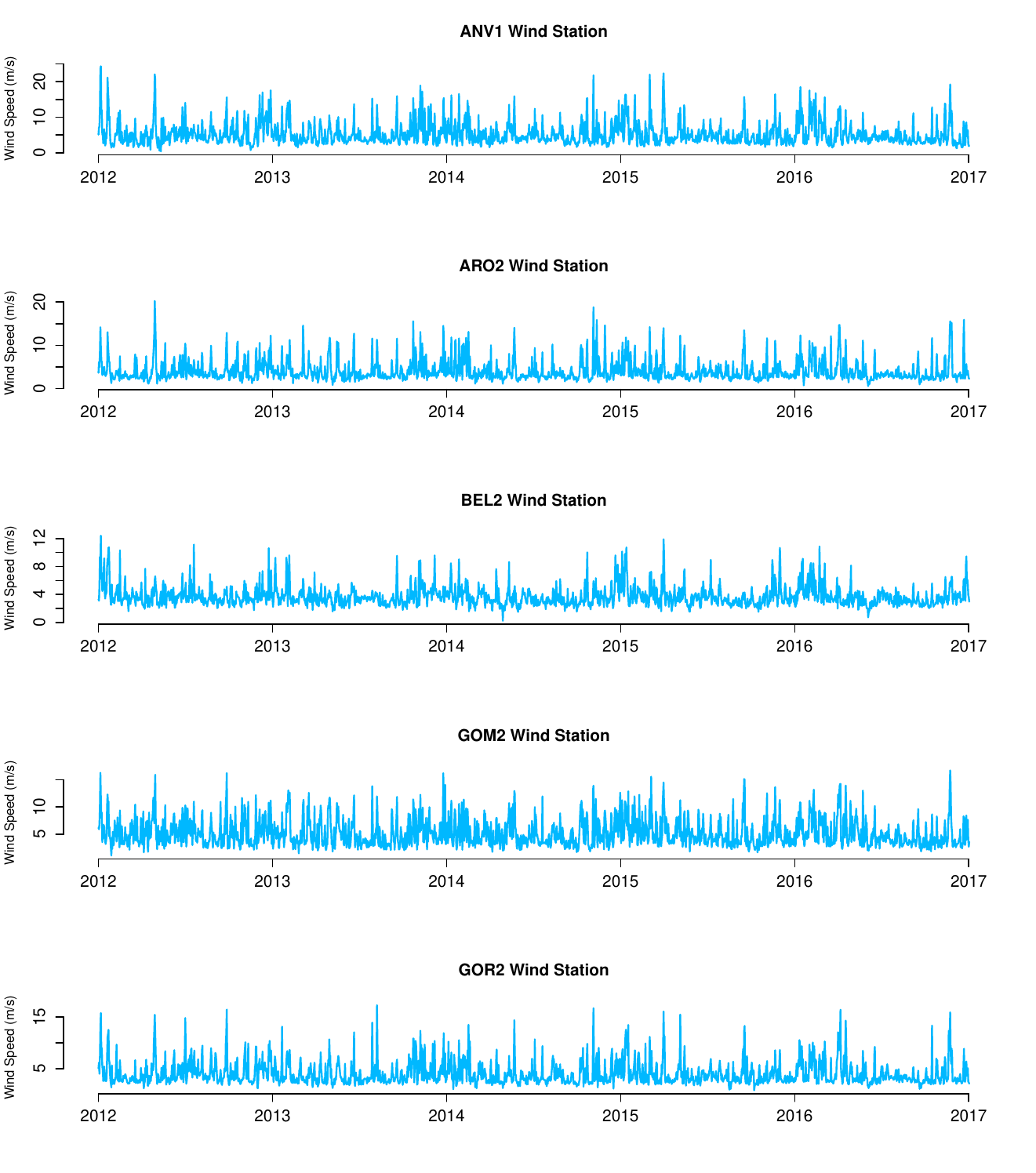}
\caption{Some wind speed time series from IMIS network. The locations of these time series are reported in Fig. \ref{fig1}. }
\label{fig2}  
\end{figure}

The data provided by MeteoSwiss were already used to investigate the multifractality of wind speed during the same investigation period  \cite{laib2018b}. Due to the presence of the 24-hour and 12-hour cycles,  the daily means of wind speed are analyzed \cite{Telesca2016}.
 Fig. \ref{fig2} shows an example of some wind speed series measured by the IMIS network. 

For all wind speed series of each network, we identified the distribution, which better fits the measured data, among the three ones that are traditionally used to model wind speed. 
 Table $1$ shows for each distribution its probability density function and the corresponding parameters. 




\begin{table}
\centering 
\begin{tabular}{p{2cm}cp{4cm}}
\hline 
Distributions  &  Density function  &  Parameters \\ 
\hline  
Weibull  &  $f(x;\lambda,k)=\left\{\begin{array}{rcl}
\frac{k}{\lambda} (\frac{x}{\lambda})^{k-1} e^{-(\frac{x}{\lambda})^{k}} \qquad x\geq 0 \\
0   \qquad \qquad  \qquad x < 0
\end{array}\right.  $  &  $k$ is the shape and $\lambda$ is  the scale \cite{reisx}.  \\
\\
Gamma & $f(x;k,\beta)=\frac{\beta^{k}x^{k-1}e^{-x\beta}}{\Gamma(k)}$  & For $x \geq0$ and $k,\beta > 0$, $k$ is the shape and $\beta$ is the rate of the Gamma distribution \cite{qpEVT}. \\
\\

Generalised Extreme Values (GEV) & $f(x;\mu, \lambda, k ) = exp\{ -[1+k(\frac{x-\mu}{\lambda})]^{\frac{-1}{k}} \} $ & where $ 1+ k(x-\mu)/\lambda > 0$, and $\mu $ is the location parameter, $\lambda$ is the scale parameter and $k$ is the shape \cite{colesB}.\\

\hline

\end{tabular}
\caption{Description of the used probability distributions.}
\label{somenoise}
\end{table} 

To this aim, we used the following Kullback-Leibler divergence (KL), a well-known quantity to evaluate the “similarity” between two distributions with density functions $p(x)$ and $q(x)$, respectively  \cite{kullback1951}:

\begin{equation}
D_{KL}(p\|q)=\int_0^\infty p(x) \log \frac{p(x)}{q(x)}dx.
\end{equation}
The reader can refer to  \cite{Ebrahimi1992, waal1996} for more details.

\begin{figure}
\centering
\includegraphics[width=\linewidth]{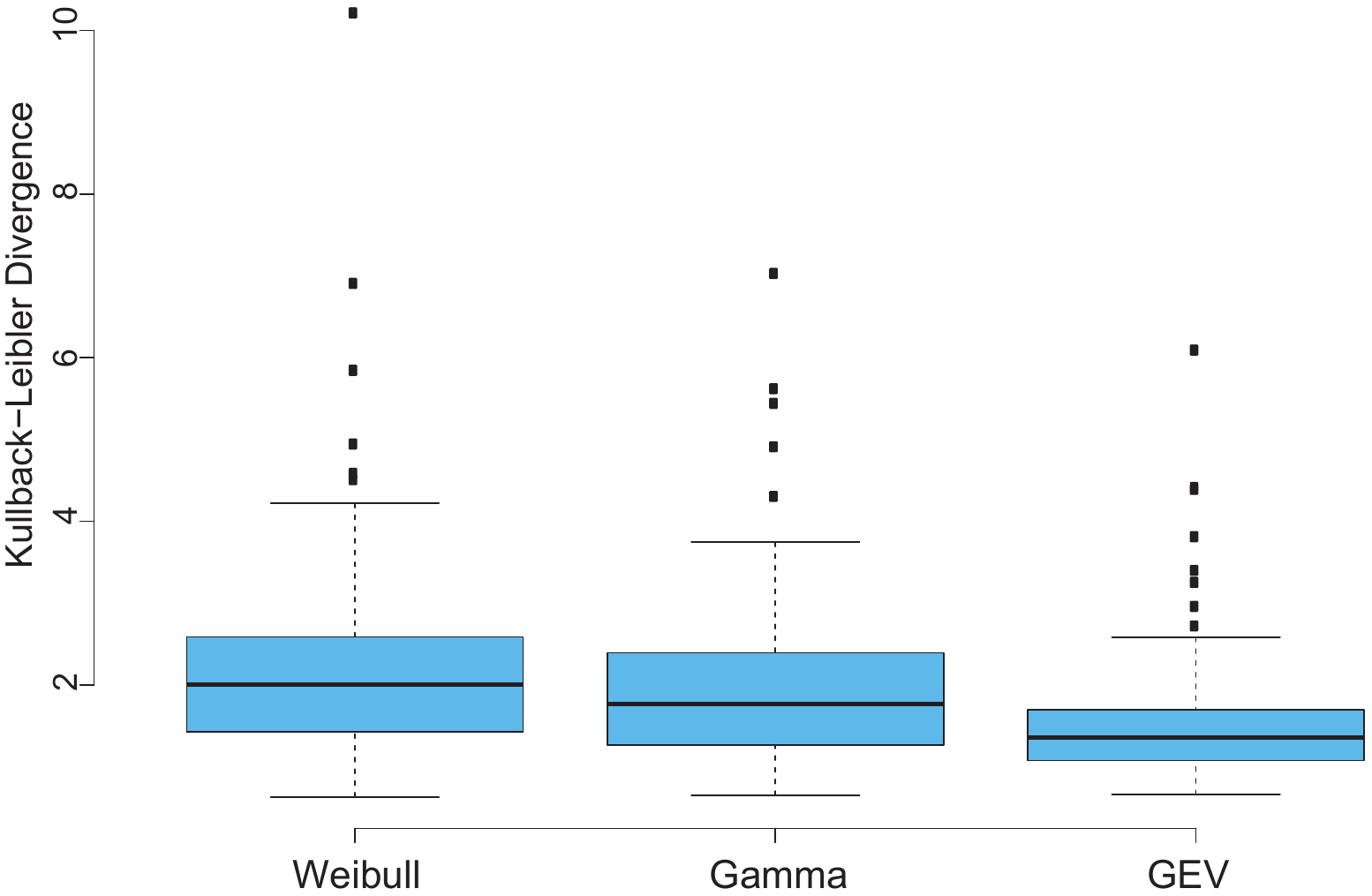}
\caption{Boxplots of the Kullback-Leibler Divergence for each probability distribution (SwissMetNet data), from \cite{laib2018b}.}
\label{fig3b}  
\end{figure}

\begin{figure}
\centering
\includegraphics[width=\linewidth]{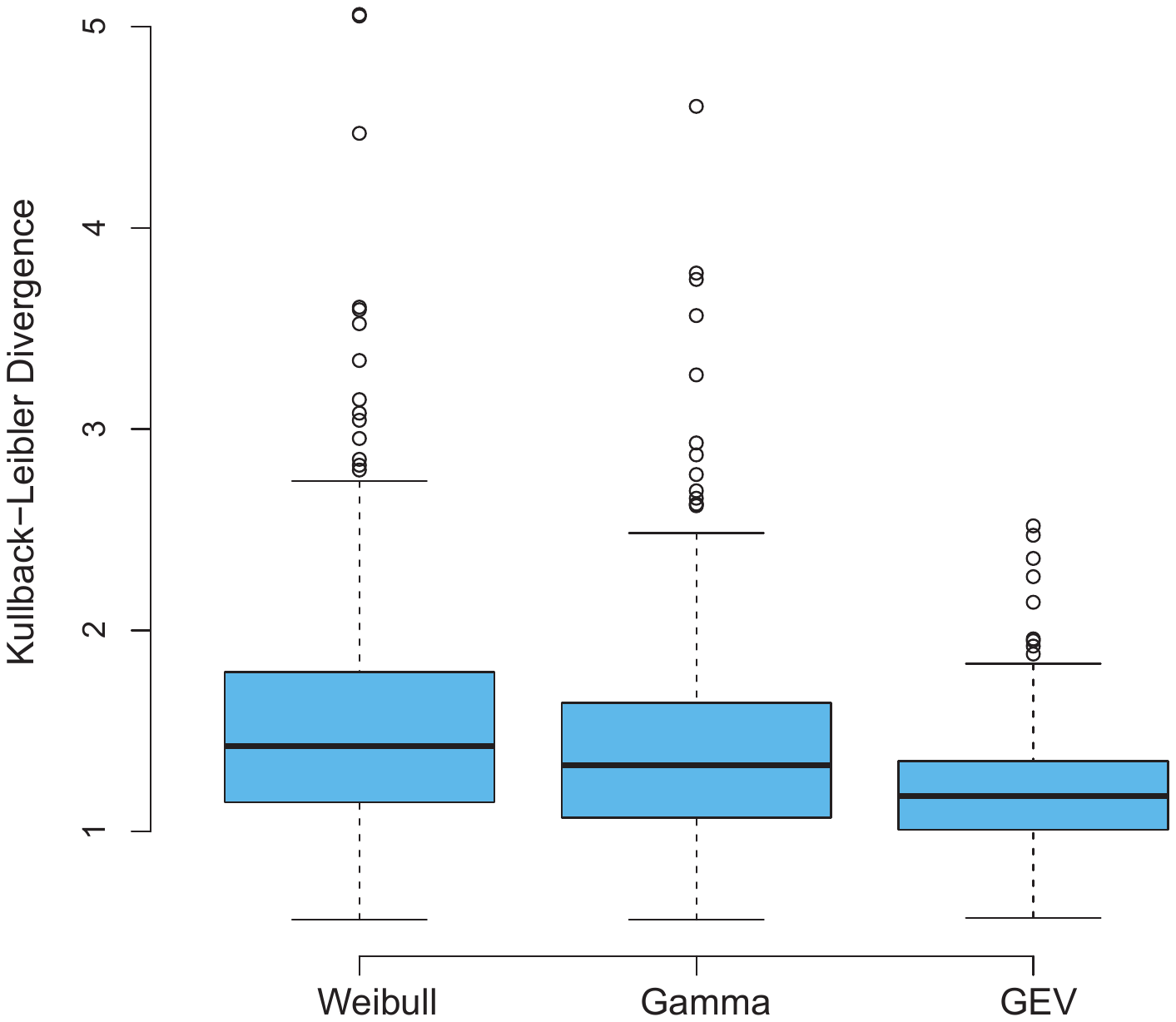}
\caption{Boxplots of the Kullback-Leibler Divergence for each probability distribution (IMIS data).}
\label{fig3}  
\end{figure}

In this paper, we just calculate the KL divergence between the proposed probability distributions (Table \ref{somenoise}) and the empirical distribution of wind speed data from the IMIS network, since the KL divergence for the SwissMetNet data was already performed in  \cite{laib2018b}.
Results from \cite{laib2018b}  are summarized in Fig. \ref{fig3b}.
The KL divergence for the IMIS network is shown in Fig. \ref{fig3}. Similarly to SwissMetNet data, the IMIS network GEV distribution seems to fit  the daily wind speed better than the other two distributions. 

\begin{figure}
\centering
\includegraphics[scale=0.5]{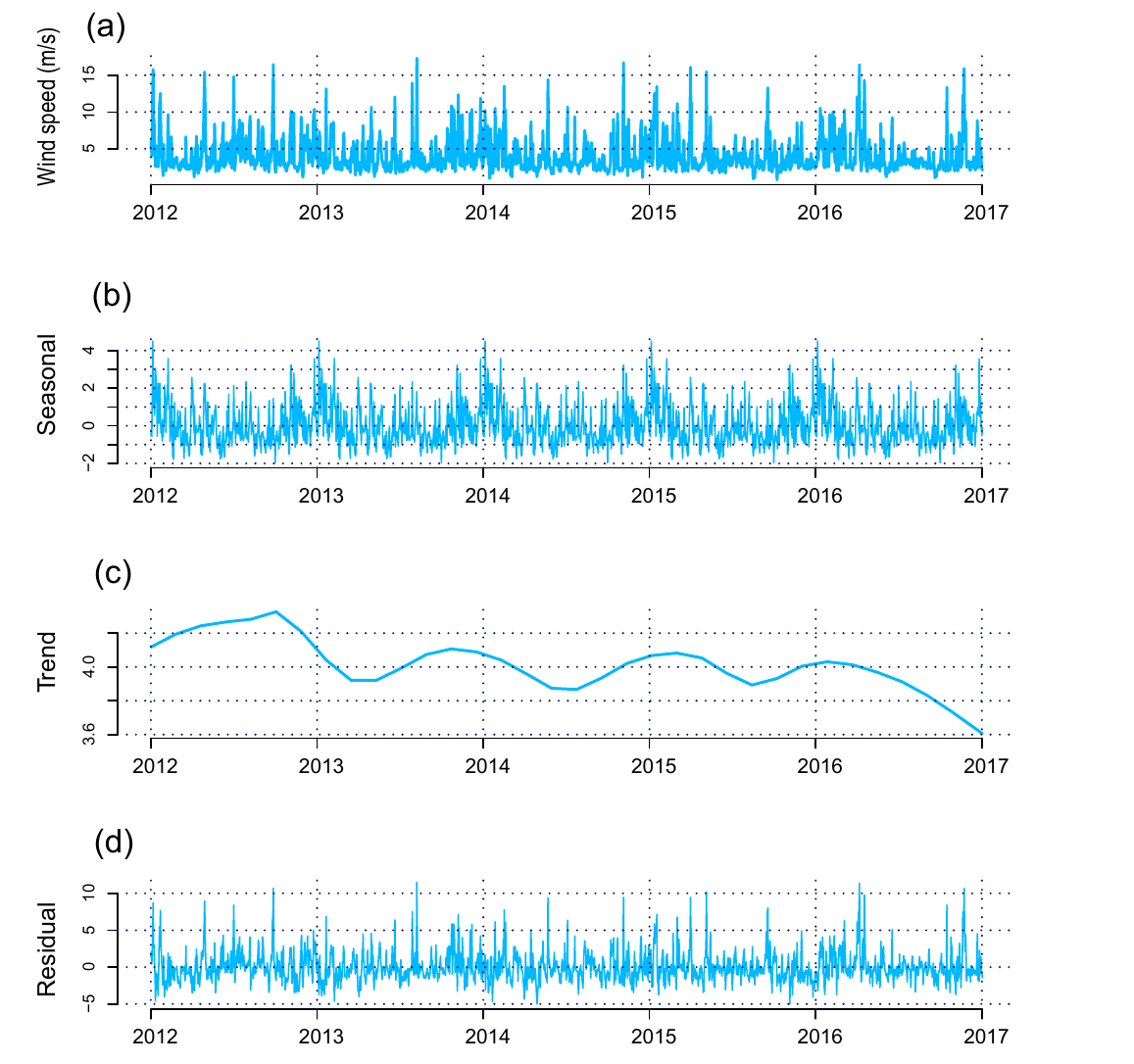}
\caption{STL decomposition of wind speed time series of IMIS station Grammont - Jumelles (GOR2). (a):  original time series, (b): seasonal component, (c): trend component, (d): remainder component.}
\label{fig4}  
\end{figure}

Due to  presence of the cycles in the wind speed, we applied to each series  the Seasonal and Trend decomposition based on the Loess smoother (STL) \cite{Cleveland1990}, which decomposed each wind speed series into trend ($T_i$), seasonal ($S_i$) and remainder ($R_i$ ) components. For our study,  only the remainder components are investigated. 
The STL decomposition, by using  the  \emph{stl} function of the \emph{"stats"} R library \cite{lanR}, was performed. The STL decomposition is known to be robust for handling extreme values and outliers. Further, each daily value of the seasonal component is calculated as the calendar mean (for instance, the value of the seasonal component at 1st January is the mean of the yearly values at 1st January of the time series).
For more theoretical details on the procedures of the STL decomposition, the reader can refer to \cite{Cleveland1990}.
Fig. \ref{fig4} shows an example of the three components for one measuring station of IMIS network. 
The seasonal component of the STL corresponds to the yearly cycle, related with the annual meteo-climatic cycle. 
The trend component is characterized by a rather slow time pattern of the wind speed with a slight yearly modulation. The remainder appears quite irregular and is characterized by residual fluctuations, which could be probably induced by the local topographic conditions at the measuring sites.


\section{Fisher-Shannon method}
The Fisher-Shannon (FS) methodology has been shown to be very efficient to analyze the complexity of the time evolution of signals. It  is based on the joint analysis of two quantities: Fisher Information Measure (FIM) \cite{fisher1925} and Shannon entropy ($H_X$) \cite{Shannon1948}. The FIM measures the amount of organization or order in a signal, while $H_X$ quantifies the amount of disorder. Fisher firstly proposed the FIM in the statistical estimation theory \cite{fisher1925}, and Frieden used it to describe the time evolution of physical systems \cite{Frieden1990}. Several applications of FIM can be found in literature. Martin et al. used this statistical quantity to gain insight into the time dynamics of electroencephalographic signals and individuate relevant changes due 
to the existence of pathological states \cite{MARTIN1999, MARTIN2001}. Lovallo and Telesca , Telesca and Lovallo  and Telesca et al. revealed the potential of  FIM to better understanding the complexity in the time dynamics of various geophysical and environmental processes \cite{Lovallo2011, Telesca2011A, TELESCA20111all, Telesca2013, Telesca2015}. Also,  it was shown that FIM could be a promising tool for recognizing precursory signs of critical phenomena in geophysical processes \cite{TELESCA2009,TELESCA2010}.

Let us indicate with $f(x)$ the density of $x$; then, its FIM $I$ is given by
\begin{equation}
I = \int^{+\infty}_{-\infty}\bigg(\frac{\partial}{\partial x}f(x)\bigg)^2 \frac{dx}{f(x)},
\end{equation}
while its Shannon entropy is defined as \cite{Shannon1948}:
\begin{equation}
H_X=-\int_{-\infty}^{+\infty}f(x)\log f(x)dx.
\end{equation}

An equivalent form of the Shannon entropy is the Shannon entropy power $N_X$, given by :
\begin{equation}
N_X=\frac{1}{2\pi e}e^{2H_X}.
\end{equation}

The calculation of the FIM and the Shannon entropy requires an estimate of the density $f(x)$ (pdf). It was recently shown in \cite{TELESCA2017} that the kernel-based approach \cite{Devroye1987, Janicki1994} to estimate the density $f(x)$ performes better than the discrete-based approach \cite{Chelani2014}. By using the kernel density estimator technique, the density $f(x)$ is approximated by
\begin{equation}
\hat{f}_M(x)=\frac{1}{Mb}\sum_{i=1}^{M}K\Bigg(\frac{x-x_i}{b}\Bigg),
\end{equation}

with $b$ the bandwidth, $M$ the length of the series and $K(u)$ the kernel function, which is a continuous non-negative function,  symmetric about zero, satisfying the following constraint

\begin{equation}
 \int_{-\infty}^{+\infty}K(u)du=1.
\end{equation}

The pdf $f(x)$ is estimated by combining the algorithms proposed in \cite{Troudi2008} and \cite{Raykar2006}, which employs a Gaussian kernel with zero mean and unit variance 
\begin{equation}
\hat{f}_M(x)=\frac{1}{M\sqrt{2\pi b^2}}\sum_{i=1}^{M}e^{-\frac{(x-x_i)^2}{2b^2}}.
\end{equation}

\section{Results}

\begin{figure}
\centering
\includegraphics[scale=0.5]{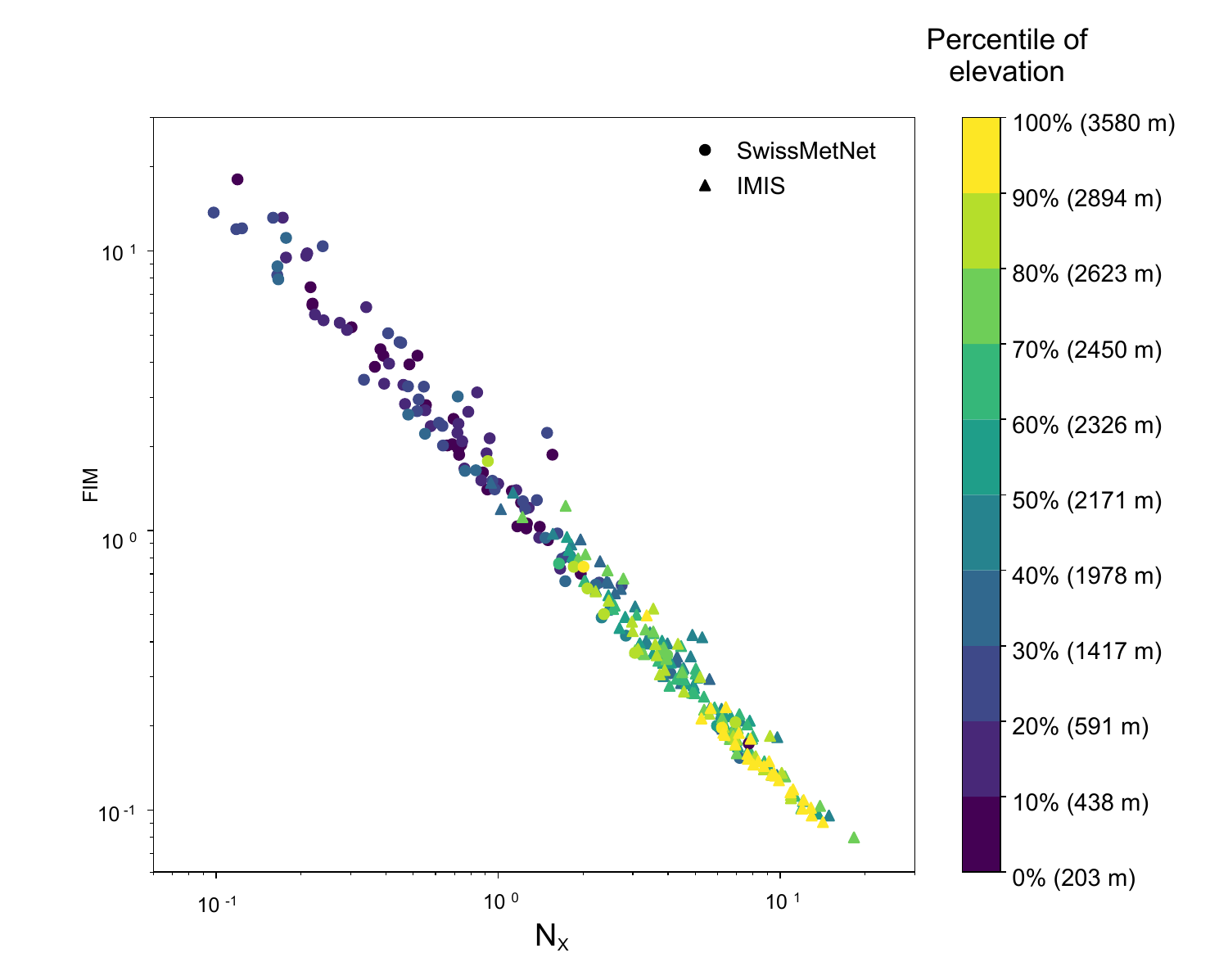}
\caption{Fisher-Shannon information plane  of the wind speed time series. The color indicates the percentile of elevation.}
\label{fig5}  
\end{figure}

\begin{figure}
\centering
\includegraphics[scale=0.5]{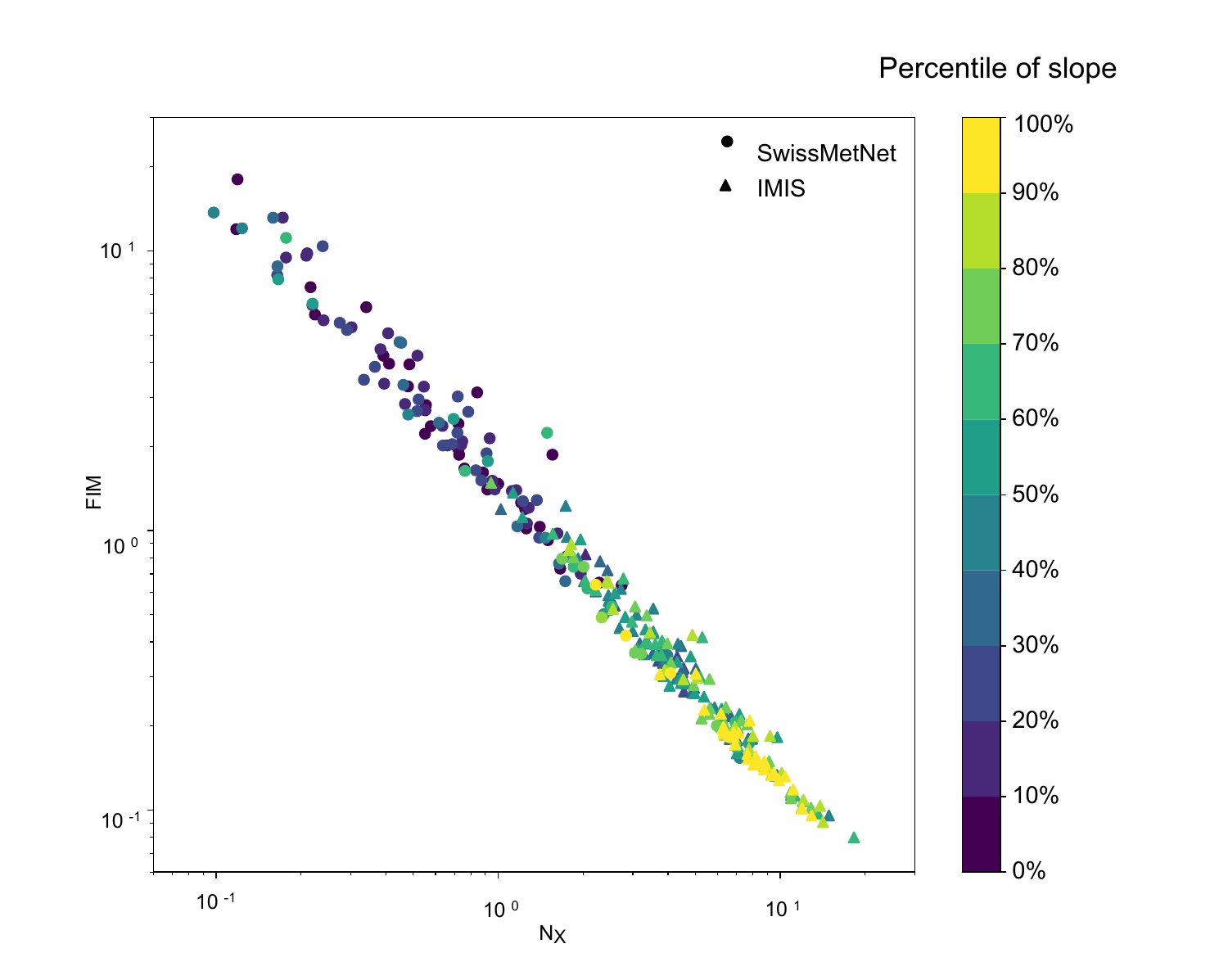}
\caption{Fisher-Shannon information plane  of the wind speed time series. The color indicates the percentile of the slope-related parameter $\mu$. }
\label{fig5bis}  
\end{figure}

For each remainder of wind speed daily means of both networks, the Fisher information measure and the Shannon entropy power were calculated.
Fig. \ref{fig5} shows the Fisher-Shannon information plane for the wind speed data, where each point represents a station (circle for the SwissMetNet network and triangle for the IMIS network), and the color of each symbol changes with the elevation of the station above sea level. Fig. \ref{fig5bis} shows the Fisher-Shannon information plane for the same wind speed data, but the color of each symbol changes with a slope-related parameter $\mu$. For each measurement site, $\mu$ was derived from a 250 m digital elevation model (Swisstopo) using the first partial derivatives of terrain elevation $z$, as shown e.g. in \cite{Helbig2017} :
\begin{equation}
\mu = \left\{ \frac{(\partial_x z)^2+(\partial_y z)^2}{2} \right\}^{1/2}.
\end{equation}
It is clearly visible that both the elevation and slope play a key role in determining the time dynamics of wind speed, which seems to be characterized by more disorder  as soon as the elevation as well as the slope increases (in fact the FIM of wind speed decreases with the elevation and slope, while the Shannon entropy power increases  with them). 

\begin{figure}
\centering
\includegraphics[width=.9\linewidth]{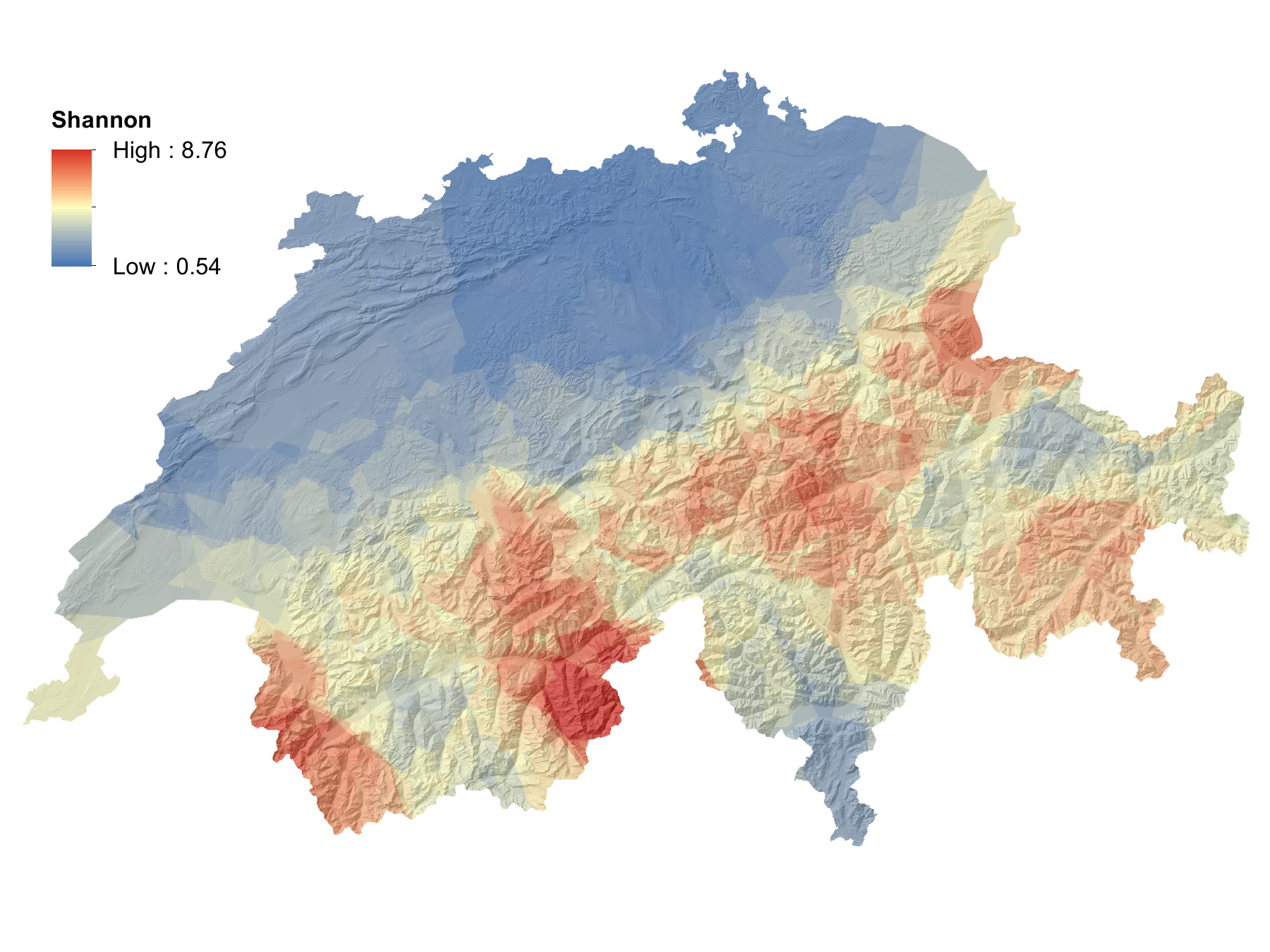}
\caption{Spatial mapping of the Shannon entropy power.}
\label{fig6}  
\end{figure}

\begin{figure}
\centering
\includegraphics[width=.9\linewidth]{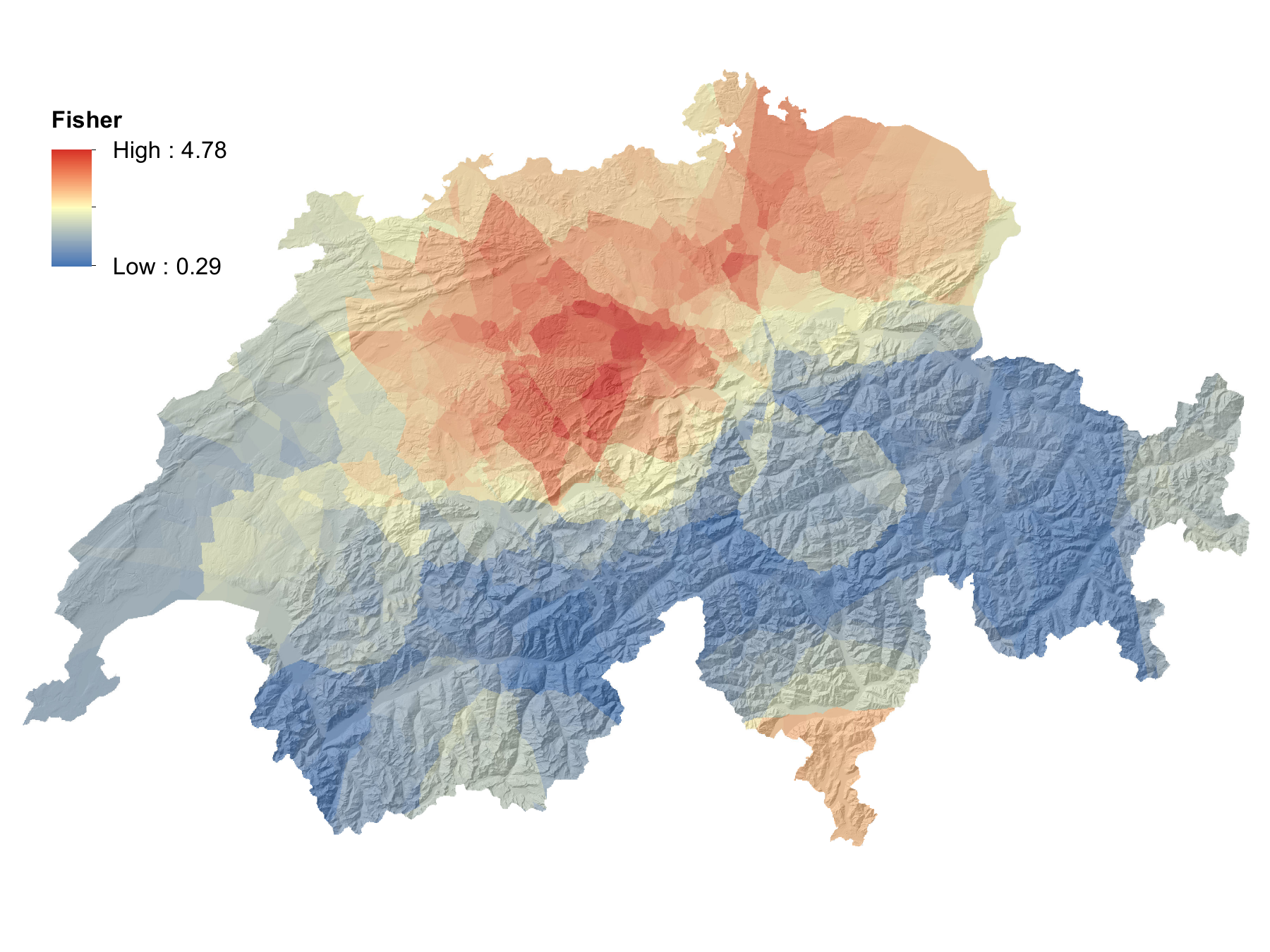}
\caption{Spatial mapping of the Fisher information measure.}
\label{fig8}  
\end{figure}

We mapped the spatial distribution of the FIM and  that of the Shannon entropy power of the wind speed on the territory of Switzerland, to identify any possible relationship with the topography. 
For this purpose, we use a well-known algorithm, called $k-$nearest neighbors ($kNN$) \cite{KPT2009}. Fig. \ref{fig6} and Fig. \ref{fig8} show the results obtained by using $kNN$, which learned the existing structure between the input (i.e. the XY-coordinates) and the output (i.e. FIM and Shannon entropy power). The maps are obtained using a 2-dimensional grid of XY coordinates (resolution of 250m) as inputs of the $kNN$ models. The prediction of the Fisher-Shannon parameters are then overlaid on the hill-shade of the digital elevation model of Switzerland.

Switzerland has three different topographic regions (Jura, Alps, and Plateau). We can see that the spatial patterns of the Fisher-Shannon parameters evidence that the largest values of the Shannon entropy power (and the lowest values of the FIM) correspond to the Alps, where the highest mountains and terrain slopes are located.

\begin{figure}
\centering
\includegraphics[width=.95\linewidth]{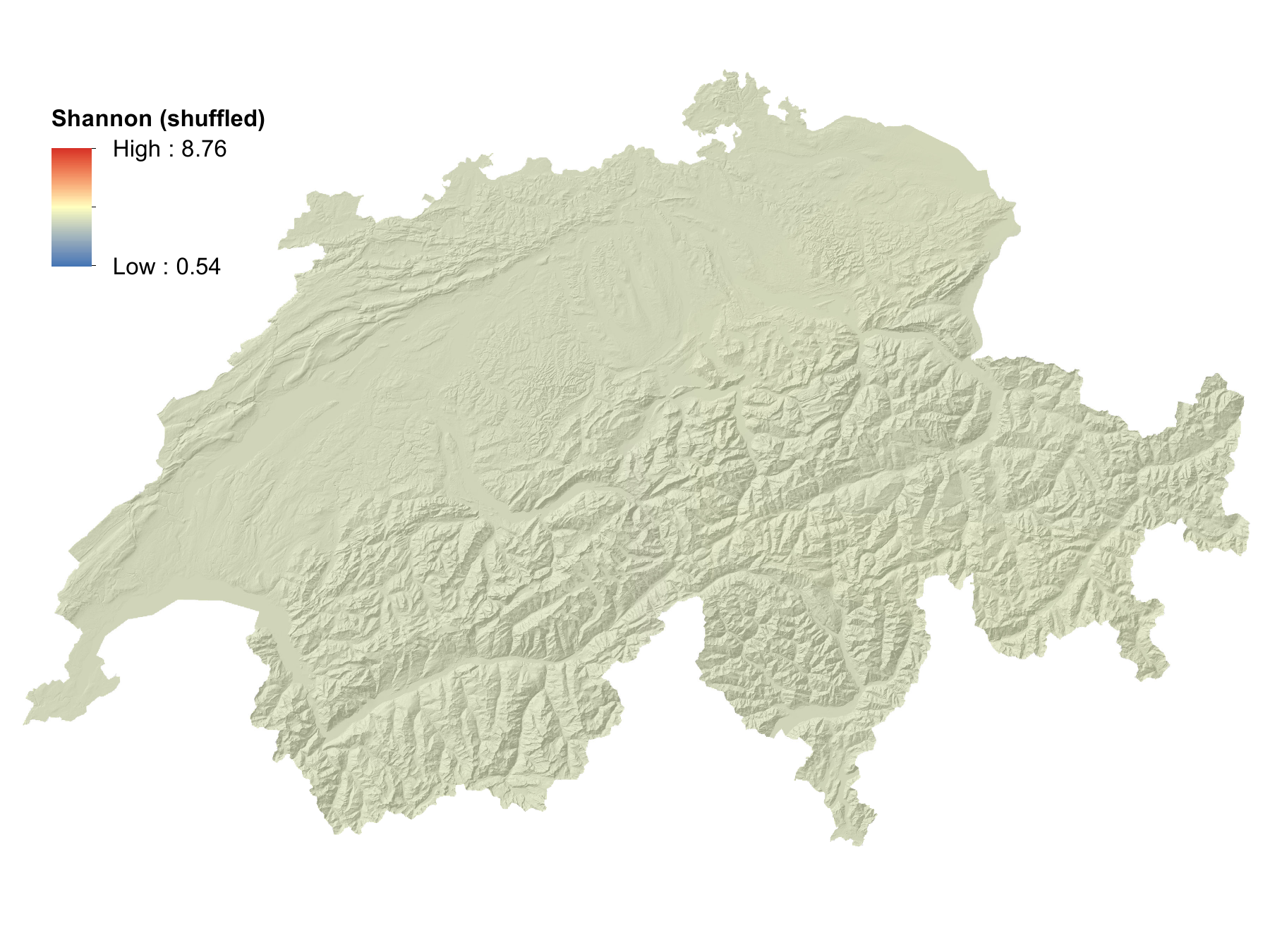}
\caption{Spatial mapping of the Shannon entropy power (shuffled stations).}
\label{fig7}  
\end{figure}

\begin{figure}
\centering
\includegraphics[width=.95\linewidth]{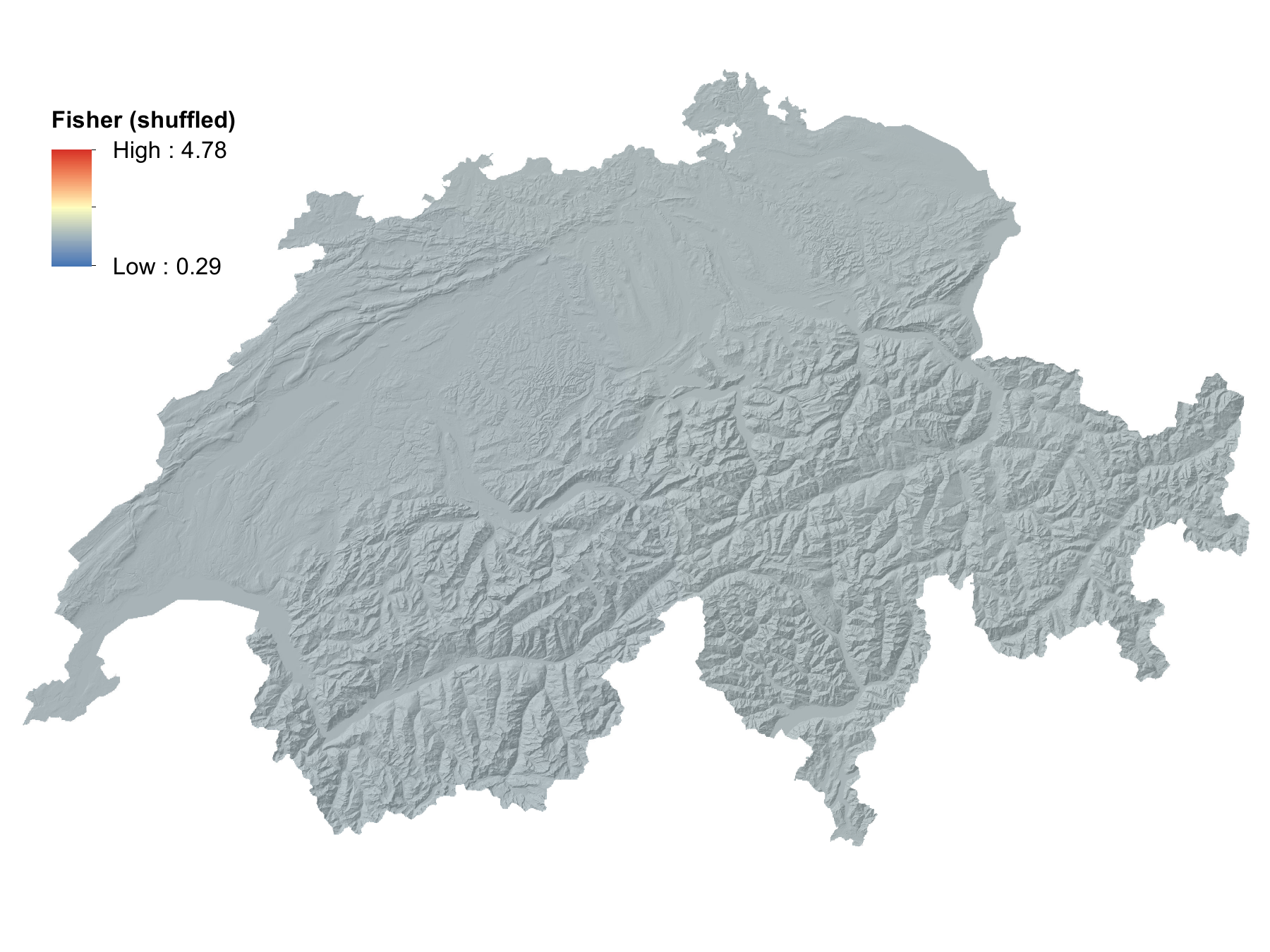}
\caption{Spatial mapping of the Fisher information measure (shuffled stations).}
\label{fig9}  
\end{figure}

In order to evaluate whether or not the obtained spatial structure was due to chance, we mapped the spatial distribution of the Fisher-Shannon parameters after shuffling the stations. Fig. \ref{fig7} and Fig. \ref{fig9} show the maps for the shuffled stations. We see that there is no more structure, which means that the results shown in  Fig. \ref{fig6} and Fig. \ref{fig8}  are not due to chance.

\section{Discussion and conclusions}
The daily means of wind speed measured at a total of 293  different monitoring stations in Switzerland were investigated in terms of their order/organization properties by using the Fisher-Shannon method. The data were preliminarily filtered  by means of the STL algorithm that decomposes a time series into trend, seasonal and remainder components. 

Such pre-processing represents an important step in our analysis, since the presence of trends and seasonal components, whose amplitude and frequency could change over time, might  bias the results obtained by applying the FS method. The STL permitted the extraction of the remainder of the wind speed, which is featured by intrinsic fluctuations that could be more probably  induced by the complex topography of the Swiss territory. 

Laib  et al. \cite{laib2018b} analyzed the multifractality of the remainder components of wind speed measured by SwissMetNet network and found that the width and the asymmetry of the multifractal spectrum (generally used to quantify the multifractality) showed a certain variability among the measuring sites, likely dependent on the morphology and complex topography of the Swiss territory.

In our study the Fisher-Shannon parameters seem to be spatially structured into two regions, one comprising the Alps and the other comprising the Plateau and the Jura range.


Although elevation represents an important factor that influences  the time dynamics of the wind speed, leading to a larger disorder for wind  speed measured at higher stations, topographic features specific to the measuring  sites should also be taken into account. In this context, a slope-related parameter used in \cite{Helbig2017} to scale spatial mean wind speed over topography correlated almost equally well with the Fisher-Shannon parameters as elevation. Terrain parameters related to slopes are commonly used to describe sheltering and exposure in mountainous terrain, influencing local wind speed patterns. We thus assume that the strong correlation with the Fisher-Shannon parameters indicates that to accurately model wind speed spatial patterns in complex topography, slope related parameters are indeed important. The Alps are characterized by very rough morphology compared with that of Jura range and the Plateau; the larger roughness could increase the randomness of the wind speed and cause the loss of order in the time dynamics.

\begin{figure}
\centering
\includegraphics[width=.95\linewidth]{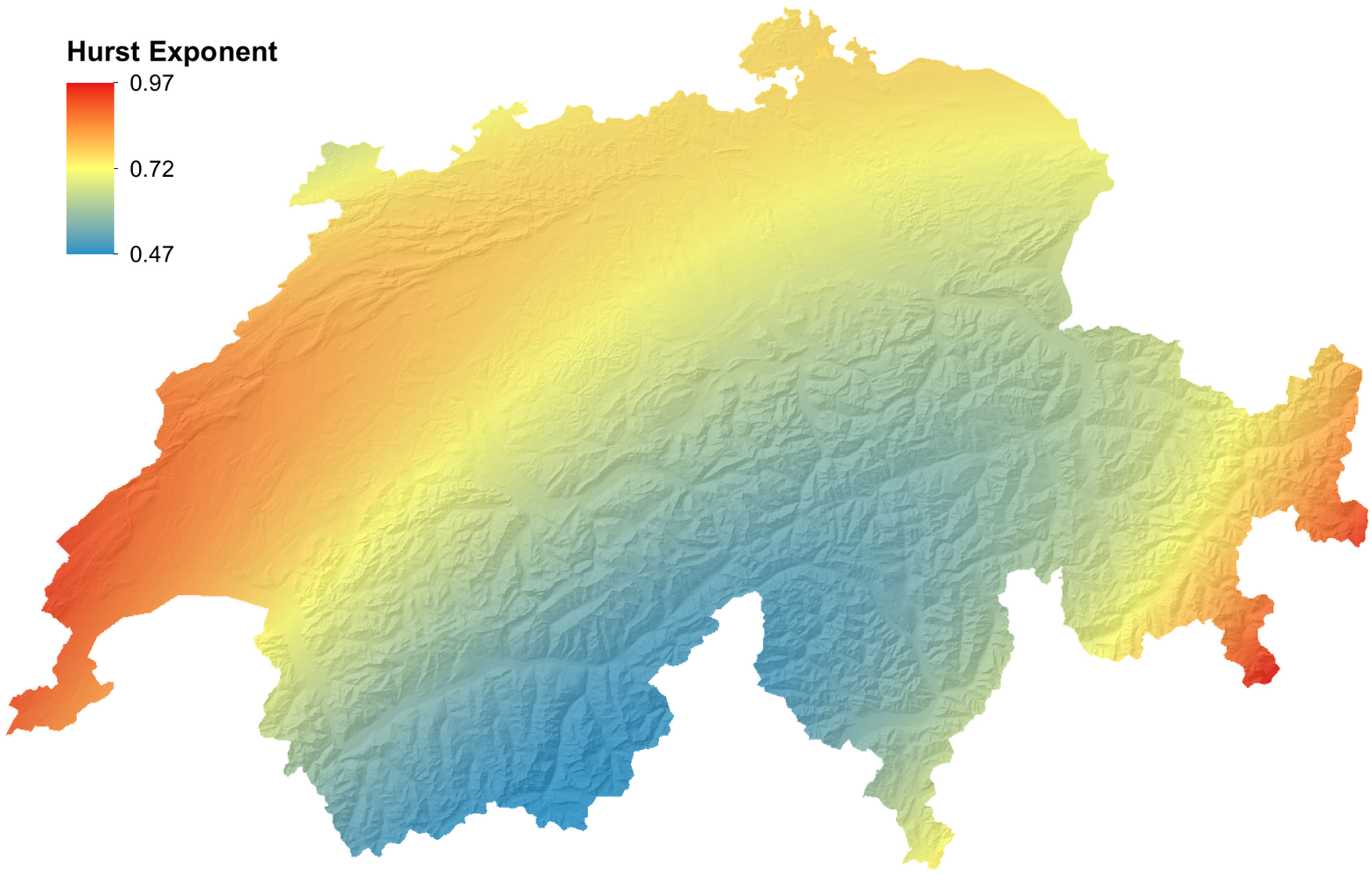}
\caption{Spatial mapping of the Hurst exponent (from \cite{laib2018b}).}
\label{fig10}  
\end{figure}

Laib et al. \cite{laib2018b} calculated for the stations of the SwissMetNet network the Hurst exponent and mapped it on the Swiss territory (Fig. \ref{fig10}). As it is clearly seen, wind speed takes on the higher values of the Hurst exponent (indicating higher persistence of the series) across the Jura range and the Plateau, while on the Alps, the Hurst exponent is lower and closer to the limit for randomness that is 0.5. The results that we have obtained by applying the FS method are, thus, consistent with those of Laib et al. \cite{laib2018b}, since the loss of order in the Alps is in a good agreement with the closeness of the wind speed series to the randomness more than shown by wind speed measured on the Jura and Plateau.

The larger order found in wind speed measured in Jura and Plateau could also be put in relationship with the peculiar morphology of the area, which “canalizes” the air blowing from the north-east to the south-west, due to the reduced distance between the Alps and the Jura Range from east to west. This makes the structure of the time dynamics of the wind speed more ordered than that featuring wind speed measured at larger elevations on the Alps, where such “canalizing” effect does not exist.

Although the present work does not completely explain  the  complexity of the temporal variability of wind speed over the whole  territory of Switzerland, it could contribute to a better characterization of wind speed and to a deeper comprehension of the mechanisms governing its dynamics.

\section{Acknowledgements}
F. Guignard and M. Kanevski thank the support of the National Research Programme 75 “Big Data” (PNR75) of the Swiss National Science Foundation (SNSF). 
M. Laib thanks the support of "Société Académique Vaudoise" (SAV) and the Swiss Government Excellence Scholarships.
L. Telesca thanks the support of the "Scientific Exchanges" project n° 180296 funded by the SNSF.

The authors thank  Alec van Herwijnen, SLF and  MeteoSwiss for providing the data. They also thank the anonymous reviewers for their constructive comments, which contributed to improving the paper.


\bibliography{xampl}
\bibliographystyle{elsarticle-num}


\newpage


%
%
%
%
%

\end{document}